\documentstyle[aps,prl,epsfig,preprint]{revtex}

\begin{document}
\tightenlines

\preprint{\parbox[b]{1in}{
\hbox{\tt PNUTP-01/A02}
}}

\draft

\title{Critical couplings in Crystalline Color superconductivity}

\author{Deog Ki Hong{\thanks{dkhong@pnu.edu}} and Y.~J. Sohn
{\thanks{mega@beauty.phys.pusan.ac.kr}}}

\vspace{0.05in}

\address{
Department of Physics, Pusan National University,
Pusan 609-735, Korea
\protect\\
\vspace{0.05in}
}


\date{\today}

\maketitle

\begin{abstract}
Solving the Schwinger-Dyson equations, we analyze the pairing of
quarks in asymmetric quark matter
where quarks have different chemical potentials.
We show that in the asymmetric quark matter a
crystalline color-superconducting gap opens
when the quark coupling is stronger than a critical value.
The critical coupling is nonzero, since the infrared divergence
is lessened when the momenta of pairing quarks are not
opposite.
The superconducting gaps and the critical couplings
are calculated both at high and
intermediate densities.
\end{abstract}


It is now widely accepted that matter at extreme density is
color superconducting quark matter~\cite{review}.
Recently, a new superconducting phase, called LOFF state,
is proposed in quark matter,
where quarks near the Fermi surface form a pair with nonzero
total momentum~\cite{Alford:2001ze,Bowers:2001ip,Leibovich:2001xr}.
The direction of the momentum is spontaneously chosen,
though its magnitude is dynamically determined. This pairing is
argued energetically more preferable, compared to BCS pairing or non-pairing,
when the difference in the chemical potentials ($\equiv2\delta\mu$)
of quarks involved in pairing lies in a certain range,
$\delta\mu_1<\delta\mu<\delta\mu_2$.

This state is originally proposed in electron superconductors
with magnetic impurities~\cite{lo,ff,takada69}.
Usually magnetic field stronger than a critical
field destroys superconductivity. But it is found
that for a certain range of magnetic field, instead of breaking Cooper pairs,
electrons form a pair with nonzero momentum, which again forms various
crystalline structures, depending on temperature and dimensionality
of the system~\cite{crystalline}.

In this letter, we analyze the superconducting gap in the LOFF phase of
dense QCD, using Schwinger-Dyson equations, to show that the LOFF pairing
occurs when the attractive force between quarks is strong
enough.
In other words, there is a critical coupling, $g_c$,
above which LOFF pairing is possible, though
$\delta\mu_1<\delta\mu<\delta\mu_2$.
This is in sharp contrast with Cooper pairing which occurs
for arbitrarily small attraction.
Since the whole Fermi surface is degenerate in Cooper pairing,
the gap equation becomes effectively $(1+1)$ dimensional and has
severe infrared divergence so that
the critical coupling for dynamical mass or gap is zero,
$g_c=0$, as one can see in the following.
Near the Fermi surface, the fermion momentum can be decomposed as
\begin{equation}
\vec p=\vec p_F+\vec l,
\end{equation}
where $\vec p_F$ is the Fermi momentum and $\vec l$ is a
residual momentum, $\left|\vec l\right|< p_F$. Since
the propagator of (free) fermions with a Fermi velocity $\vec v_F$
is given near the Fermi surface as
\begin{equation}
S_F(l_0,\vec l)={i\over (1+i\epsilon)l_o-\vec v_F\cdot \vec l},
\end{equation}
the excitation energy does not depend on the
perpendicular momentum
$\vec l_{\perp}\equiv \vec l-\vec v_F\vec l\cdot\vec v_F$.
For Cooper pairing, the gap depends only on
$l_{\parallel}=(l_0,\vec v_F\vec v_F\cdot\vec l)$, since the pairing
quarks have opposite momenta and the whole Fermi surface is degenerate.
The Cooper-pair gap equation (in Euclidean space)
therefore becomes (1+1) dimensional,
if integrated over the perpendicular momentum $\vec l_{\perp}$:
\begin{equation}
\Delta(p_{\parallel})=g^2\int {{\rm d}^4l\over (2\pi)^4}
G(p_{\parallel}-l) {\Delta(l_{\parallel})\over
l_{\parallel}^2+\Delta^2}=
g^2\int {{\rm d}^2l_{\parallel}\over
(2\pi)^2}{\tilde G}(p_{\parallel}-l_{\parallel})
{\Delta(l_{\parallel})\over l_{\parallel}^2+\Delta^2}~,
\end{equation}
where $G$ is the kernel for Cooper pairing.
For generic kernels, the Cooper-pair gap equation has a nontrivial
solution for any coupling $g^2>0$:
$\Delta\sim \exp(-c/g^2)$ for four-Fermi interactions or for
a constant kernel~\cite{fourFermi} and $\Delta\sim \exp(-c^{\prime}/g)$
for Landau-damped gluons~\cite{Son:1999uk,Hong:2000tn,landau}.

Another way of seeing the instability of Fermi surface against
formation of Cooper pairs for any arbitrarily weak attraction
is by the renormaliztion group (RG) analysis.
When incoming fermions have opposite momenta but equal in
magnitude, the four-Fermi interaction is marginally
relevant and thus develops Landau pole as we scale down to the
Fermi surface~\cite{Polchinski:1992ed}.
In Cooper pairing, it is therefore crucial that the pairing
fermions have opposite momenta, lying near the Fermi surface.

On the other hand, when the mismatch between two
Fermi surfaces is large enough, $\delta\mu>\Delta/\sqrt{2}$,
it is energetically not favorable to form Cooper pairs since
at least either one of the pairing fermions has to be excited
far from the Fermi surface, costing too much energy in
pairing~\cite{Alford:1999pa,Alford:2001ze}.
In the LOFF phase, however,
both of the fermions involved in pairing may lie near the Fermi surface
by forming a pair of nonvanishing momentum, costing much less energy.
(See Fig.~1.) Though the momenta of pairing fermions are opposite
in the rest frame of the pair (Fig.~1 (b)),
they are not in the rest frame of Fermi sea or in the
ground state (Fig.~1 (a)).
For such pairing, the (effective) four-Fermi interaction
is not marginal and thus does not lead to Landau pole or
dynamical mass unless the interaction is strong
enough, which is a characteristic feature
in dimensions higher than (1+1)~\cite{critical,qed}.
As we will see later, LOFF pairing indeed occurs
in dense QCD with light flavors when
the couplings are bigger than critical values for both high and
intermediate density.


When the chemical potentials of up and down quarks in quark matter
are different, $\mu_d-\mu_u(\equiv2\delta\mu)\ne0$,
LOFF pairing of quarks may be possible
for a certain range of $\delta\mu\in\left[\delta\mu_1,\delta\mu_2\right]$.
The ground state is then described by a condensate of LOFF pairs,
\begin{equation}
\left<\psi_d^i(\vec p+\vec q)\psi_u^j(-\vec p+\vec q)\right>=
\epsilon^{ij3}\Delta(\vec q),
\end{equation}
where $i,j=1,2,3$ are color indices and the unbroken color direction
is denoted as 3.
For a given $\vec q$, the momenta of the quarks involved in LOFF pairing
can be decomposed as
\begin{equation}
p+q=\mu_d v_d+l,\quad q-p=\mu_uv_u+l^{\prime},
\end{equation}
where $v^{\mu}=(0,\vec v_F)$, $q^{\mu}=(0,\vec q)$,
and the residual momenta are restricted as
$\left|l^{\mu}\right|<\mu_d$ and $\left|l^{\prime\mu}\right|<\mu_u$.
$\vec v_F^d$ and $\vec v_F^u$ are Fermi
velocities of the down and up quarks, respectively.
As shown in the effective theory developed in~\cite{Hong:2000tn},
the field that
describes the quarks near the Fermi surface is given as
\begin{equation}
\psi(\vec v_F,x)={1+\vec\alpha\cdot \vec v_F\over2}
e^{i\mu\vec v_F\cdot\vec x}\psi(x),
\end{equation}
where $\vec\alpha=\gamma_0\vec \gamma$ and
$\psi(x)$ is the quark field that contains both negative and
positive energy states. Then, the LOFF condensate in the position space
can be written as
\begin{equation}
\left<\psi_d^i(\vec v_F^d,x)\psi_u^j(\vec v_F^u,x)\right>
=\epsilon^{ij3}K(\vec q),
\end{equation}
where the Fermi velocities satisfy
$\mu_d\vec v_F^d+\mu_u\vec v_F^u=2\vec q$.
The magnitude of the LOFF-pair momentum is determined
by minimizing the ground state energy, which is
a function of $\delta\mu$ and the coupling $g$.
But, in this letter we study
the LOFF gap, assuming $\vec q$ is given.

Once a LOFF gap opens, quarks get a Fermi-momentum dependent
Majorana mass, off-diagonal in the basis of Fermi velocities, given as
\begin{equation}
{\cal L}_{\rm mass}=-\sum_{\vec v_F^u}
\epsilon^{ij3}\Delta^{\dagger}(\vec q)\psi_d^i(\vec v_F^d,x)
\psi_u^j(\vec v_F^u,x)+h.c.
\end{equation}
with $\mu_d\vec v_F^d+\mu_u\vec v_F^u=2\vec q$.
Now, let us introduce Nambu-Gorkov fields to describe the LOFF gap,
\begin{equation}
\Psi(\vec v_F,x)=\pmatrix{\psi_d(\vec v_F^d,x) \cr
\psi_u^c(-\vec v_F^u,x)\cr},
\end{equation}
where the charge-conjugate field is defined as
$\psi_u^c(-\vec v_F,x)=C\bar\psi_u^T(+\vec v_F,x)$
with $C=i\gamma^0\gamma^2$.
The inverse propagator of the Nambu-Gorkov field is then
given as
\begin{equation}
S^{-1}(\vec v_F,l)= -i\gamma_0\pmatrix{Z(l)l\cdot V_d &
      -\Delta(\vec q,l) \cr
-\Delta^{\dagger}(\vec q,l)& Z(l)l\cdot V_u\cr},
\end{equation}
where $Z(l)$ is the wave function renormalization constant
and $\Delta(\vec q,l)$ is the LOFF pair gap, and we have
introduced $V^{\mu}=(1,\vec v_F)$.

We first analyze the gap equations for quark matter at intermediate
density where QCD interactions are described by effective
four-Fermi interactions with a ultraviolet cutoff $\Lambda$.
If we take the four-Fermi interaction to mimic the one-gluon exchange
interaction, the gap equation is given as
\begin{equation}
\Delta(\vec q)=-i{2\over3}{g^2\over \Lambda^2}\int{{\rm d}^4k\over (2\pi)^4}
{\Delta(\vec q)\over k\cdot V_d k\cdot V_u-\Delta^2},
\label{gap}
\end{equation}
where the factor $2/3$ is inserted since the gap is color antitriplet.
The characteristic feature of the gap equation (\ref{gap})
for the LOFF paring
is that the quark propagator
is a function of three independent momenta, $k_0$,
$\vec k\cdot \vec v_F^u(\equiv k_u)$,
and $\vec k\cdot\vec v_F^d(\equiv k_d)$,
while in BCS pairing it is a function of two,
$k_0$ and $\vec k\cdot\vec v_F$.
In general, we may decompose a momentum $\vec k$ as
\begin{equation}
\vec k=\vec k_{\perp}+k_u\vec v_u^*+k_d\vec v_d^*,
\end{equation}
where $\vec v_{u,d}^*$ are dual to $\vec v_F^{u,d}$,
satisfying
$\vec v_F^a\cdot\vec v_b^*=\delta^a_b$ with $a,b=u,d$.

Since the quark propagator is independent of $\vec k_{\perp}$, it just
labels the degeneracy in the LOFF pairing.
The perpendicular momentum $\vec k_{\perp}$ forms a circle on the Fermi
surface, whose radius is given as
$\mu_d\sin\alpha_d~(=\mu_u\sin\alpha_u)$, where $\alpha_{d,u}$ are the angles
between $\vec q$ and $\vec v_F^{d,u}$, respectively.
Upon integrating over $\vec k_{\perp}$, the gap equation (\ref{gap})
becomes a (2+1) dimensional gap equation. This is in sharp contrast with
the gap equation in the BCS pairing, which is (1+1) dimensionl after
integrating over the $\vec k_{\perp}$, namely over the whole Fermi surface.

Integrating over $\vec k_{\perp}$, we find
the gap equation in Eucleadian space to be
\begin{eqnarray}
1&=&{2g^2\mu_d\sin\alpha_d\over3 \Lambda^2\sin\beta}
\int{{\rm d}k_0\over2\pi}{{\rm d}k_u{\rm d}k_d\over(2\pi)^2}
{1\over (k_0-ik_d)(k_0-ik_u)+\Delta^2}
\label{gap_loff}\\
&\simeq &{2g^2\mu_d\sin\alpha_d\over3 \Lambda^2\sin\beta}
\int_{\Delta}^{\Lambda}{{\rm d}k_0\over2\pi^3}\left[
\ln\left({\Lambda\over k_0}\right)\right]^2
\end{eqnarray}
where $(\sin\beta)^{-1}$ is the Jacobian and $\beta$ is the angle
between $\vec v_F^d$ and $-\vec v_F^u$.
In integrating over $k_u$ and $k_d$ in Eq.~(\ref{gap_loff}),
we have restricted $k_u$ and $k_d$ to have opposite sign,
since we are interested either in the quark pair or
in the hole pair (but not in the quark-hole pair).
Finally integrating over $k_0$, we get
\begin{equation}
1-{g_c^2\over g^2}={1\over2}\left({\Delta\over\Lambda}\right)
\left[\ln\left({\Delta\over\Lambda}\right)\right]^2,
\end{equation}
where
\begin{equation}
g_c^2={3\pi^3\Lambda\sin\beta\over 4\mu_d\sin\alpha_d}.
\end{equation}
Therefore, we see that the (LOFF) gap opens only if the four-Fermi coupling
$g^2>g_c^2$, as we have expected.
If we take $\mu=400~{\rm MeV}$ and $\Lambda=200~{\rm MeV}$,
the critical coupling for the LOFF gap at the intermediate
density is $g_c^2\simeq11.6\sin\beta\,(\sin\alpha_d)^{-1}$.
The size of the gap is determined by how close the coupling
is to the critical coupling $g_c^2$. Near the critical coupling,
we find the gap to be
\begin{equation}
\Delta\simeq 2\Lambda {1-g_c^2/g^2\over
\left[\ln\left(1-g_c^2/g^2\right)\right]^2}.
\end{equation}

Now, we analyze the gap equation for high density quark matter, where
the magnetic gluons are not screened but Landau damped.
In the hard-dense loop (HDL)
approximation, the Schwinger-Dyson (SD) equation
becomes in the leading order, taking $Z=1$,
\begin{equation}
\Delta({\vec q},l)=\left(-ig_s\right)^2
\int{{\rm d}^4k\over (2\pi)^4}V_u^{\mu}D_{\mu\nu}(l-k)V^{\nu}_d
{T^a\Delta({\vec q},k)T^a\over k\cdot V_dk\cdot V_u-\Delta^2({\vec q},k)},
\label{sd}
\end{equation}
where $T^a$ is the color generator in the fundamental representation
and $D_{\mu\nu}$ is the gluon propagator, given in the HDL approximation
as
\begin{equation}
iD_{\mu\nu}(k)={P_{\mu\nu}^{T}\over k^2-G}+{P^L_{\mu\nu}\over k^2-F}
-\xi {k_{\mu}k_{\nu}\over k^4},
\end{equation}
where $\xi$ is the gauge parameter and
the projectors are defined by
\begin{eqnarray}
P^T_{ij}&=&\delta_{ij}-{k_ik_j\over |\vec k|^2},
\quad P_{00}^T=0=P_{0i}^T\\
P^L_{\mu\nu}&=&-g_{\mu\nu}+{k_{\mu}k_{\nu}\over k^2}-P^T_{\mu\nu}.
\end{eqnarray}
The vacuum polarization functions in medium becomes
in the weak coupling limit ($|k_0|\ll|\vec k|$) as
\begin{eqnarray}
F(k_0,\vec k)\simeq M^2,\quad
G(k_0,\vec k)\simeq {\pi\over 4}M^2{k_0\over |\vec k|},
\end{eqnarray}
where the screening mass
$M^2=g_s^2{\bar\mu}^2/\pi^2$ with $2\bar\mu^2=\mu_u^2+\mu_d^2$.

Since the main contribution to the integration in Eq.~(\ref{sd})
comes from the momenta in the region
\begin{equation}
k_0,k_u,k_d \sim\Delta \quad {\rm and} \quad
\left|\vec k_{\perp}\right|\sim M^{2/3}\Delta^{1/3},
\end{equation}
the leading contribution is by the Landau-damped magnetic gluons.
For such momentum range,
\begin{equation}
V_u\cdot P^T\cdot V_d\simeq -\vec v_F^u\cdot \vec v_F^d\equiv \cos\beta(>0),
\end{equation}
where we define the angle between $\vec v_F^d$ and
$-\vec v_F^u$ as $\beta$.
From the gap equation, we note that at the leading order
the LOFF gap is a function of energy only:
$\Delta(\vec q,l)\simeq \Delta(\vec q,l_0)$.
Therefore, at the leading order the gap equation becomes
in Euclidean space as
\begin{equation}
\Delta(\vec q,l_0)={2\over3}g_s^2\cos\beta
\int{{\rm d}^4k\over (2\pi)^4}{1\over {\vec k}^2+{\pi\over4}M^2
|k_0-l_0|/|\vec k|}\cdot{\Delta(\vec q,k_0)
\over k\cdot V_E^u k\cdot V_E^d+\Delta^2},
\label{gap1}
\end{equation}
where the factor $2/3$ is due to the color factor
and $V_E^{\mu}\equiv(1,-i\vec v_F)$.
Decomposing the loop momentum into $\vec k=\vec k_{\perp}
+k_u\vec v_u^*+k_d\vec v_d^*$ with the Jacobian $(\sin\beta)^{-1}$,
we integrate over $\vec k_{\perp}$ to find at the leading order
in the $1/\bar\mu$ expansion
\begin{eqnarray}
\Delta(\vec q,l_0)&\simeq&{2g_s^2\cot\beta\over9\sqrt{3}}
\left({4\over\pi M^2}\right)^{1/3}
\int{{\rm d}k_0\over2\pi}
{{\rm d}k_u{\rm d}k_d\over(2\pi)^2}{1\over |k_0-l_0|^{1/3}}
{\Delta(\vec q,k_0)\over (k_0-ik_u)(k_0-ik_d)+\Delta^2},
\label{cgap}
\end{eqnarray}
where the integration ranges are taken to be
$\left|k_{u,d}\right|\le\bar\mu$ and $\left|k_0\right|\le \bar\Lambda$.
To convert the gap equation into a differential equation, we approximate
the kernel as
\begin{equation}
\left|l_0-k_0\right|^{-1/3}\simeq
\left\{
\begin{array}{ll}
\left|l_0\right|^{-1/3} & \mbox{if $\left|l_0\right|>
\left| k_0\right|$},\\
\left|k_0\right|^{-1/3} & \mbox{otherwise}.
\end{array}
\right.
\end{equation}
Then, the gap equation becomes
\begin{equation}
x^{4/3}{{\rm d}^2y\over {\rm d}x^2}+
{4\over3}x^{1/3}{{\rm d}y\over {\rm d}x}+
r y F(x, \Delta,\bar\mu)=0, \quad {\rm with }\quad
r={2g_s^{4/3}\cot\beta\over 9\sqrt{3}\pi^2(4\pi)^{2/3}}
\end{equation}
where $x\equiv l_0/\bar\mu$, $y\equiv\Delta(\vec q,l_0)/\bar\Lambda$
and
\begin{equation}
F(x,\Delta,\bar\mu)=
\int{{\rm d}k_u{\rm d}k_d
\over (k_0-ik_u)(k_0-ik_d)+\Delta^2}.
\end{equation}
For $l_0\ll\bar\mu$ or $x\ll 1$,
the equation becomes
\begin{equation}
x^{4/3}{{\rm d}^2y\over {\rm d}x^2}+
{4\over3}x^{1/3}{{\rm d}y\over {\rm d}x}+
r y\left[\ln\left({\bar\mu
\over\Delta}\right)\right]^2=0
\end{equation}
and the solution is found to be
\begin{equation}
y\simeq {A\over x^{1/6}}\, J_{1/2}\left(3\sqrt{r}x^{1/3}
\ln\left[\bar\mu/\Delta\right]\right),
\end{equation}
where $J$ is the Bessel function and
$A=\pi\Delta(0)\left[6\sqrt{r}\ln(\bar\mu/\Delta)
\right]^{-1/2}/\bar\Lambda$.
Similarly for $x\gg1$ or $l_0\gg\bar\mu$,
\begin{equation}
x^{4/3}{{\rm d}^2y\over {\rm d}x^2}+
{4\over3}x^{1/3}{{\rm d}y\over {\rm d}x}+
r x^{-2}y=0.
\end{equation}
We find the solution for large $x$
\begin{equation}
y\simeq {B\over x^{1/6}}J_{1/4}\left({3\sqrt{3}\over2}x^{-2/3}\right)
\sim x^{-1/3},
\end{equation}
where $B$ is a constant.
We see that the gap vanishes rather slowly as $l_0$
approaches $\bar\Lambda$.

Now, let us find the critical coupling at which the gap vanishes.
Taking $\Delta=0$ in the gap equation (\ref{gap1}), we get
\begin{equation}
1={2\over3}g_c^2\cos\beta
\int
{{\rm d}^4k\over (2\pi)^4}{1\over {\vec k}^2+{\pi\over4}M^2
|k_0|/|\vec k|}\cdot{1\over k\cdot V_E^u k\cdot V_E^d}.
\end{equation}
Integrating over $\vec k$ with the restriction that
$k_u$ and $k_d$ should have an opposite sign, we get
\begin{eqnarray}
1
&\simeq&{2g_c^{4/3}\cot\beta\over 9\sqrt{3}\pi^2}
\left({4\pi\over {\bar\mu}^2}\right)^{1/3}
\int_0^{\bar\mu}{{\rm d}k_0\over 2\pi}{1\over |k_0|^{1\over3}}
\left[\ln\left({\bar\mu\over k_0}\right)\right]^2={\sqrt{3}g_c^{4/3}
\cot\beta \over \pi^2(4\pi)^{2/3}}.
\end{eqnarray}
Therefore, we find
the critical coupling is $g_c\simeq 13\, (\tan\beta)^{3/4}$.
For a rough estimate of the gap, we may take
$\Delta(\vec q,k_0)\simeq\Delta(\vec q)$ for all $k_0<\bar\Lambda$
in Eq.~(\ref{cgap}) and find
\begin{equation}
1-\left({g_c\over g_s}\right)^{4/3}\simeq
{2\over9}\left({\Delta\over\bar\Lambda}\right)^{2/3}
\left[\ln\left({\bar\Lambda\over\Delta}\right)
\right]^2,
\label{hgap}
\end{equation}
which shows that the LOFF gap opens only when the strong
coupling $g_s$ is stronger than a critical coupling,
$g_c\simeq 13\, (\tan\beta)^{3/4}$.
We see that at high density where the coupling is weak the LOFF
gap opens when the momenta of pairing quarks are almost
opposite or when $\vec q$ is quite small.
Near the critical coupling, we solve Eq.~(\ref{hgap}) to find
\begin{equation}
\Delta\simeq {2\sqrt{2}\bar\Lambda\over 27g_s^2}\,
{\left(g_s^{4/3}-g_c^{4/3}\right)^{3/2}
\over \left|\ln\left(g_s^{4/3}-g_c^{4/3}\right)\right|^3}.
\end{equation}

In conclusion, we show that the LOFF gap opens in asymmetric quark matter
when the quark coupling
is stronger than a critical value, since the gap equation is much less
divergent in the infrared region, compared to the Cooper pair
gap equation. Both the critical couplings and the gaps for the
LOFF phase are calculated at intermediate and at high density.
We find that the critical couplings become smaller as the angle
between the momenta of
pairing quarks approaches $\pi$. The angle is determined, once
the pairing momentum is obtained by
minimizing the ground state energy.
This may explain why it is very hard to realize
the crystalline superconductivity in ordinary BCS superconductors,
where the electron
couplings are quite small. On the other hand, it is quite possible to
observe the crystalline superconductivity in High $T_C$ superconductors
or heavy fermion superconductors~\cite{gloos}
where the effective electron coupling is large and there is no
$\mu/\Lambda$ enhancement due to the degeneracy in the gap equation for
two dimensional systems. Therefore,
it will be quite interesting to find a condensed system where one can test
the coupling dependence of the crystalline superconducting gap
obtained in this letter.



\acknowledgments

We are grateful to J. Bowers, C. Kim, K. Rajagopal, and D. Seo
for useful discussions.
One of us (D.K.H.) wishes to thank
the Center for Theoretical Physics, MIT,
where part of this work was done, for its hospitality.
This work was supported by Korea Research Foundation Grant
(KRF-2000-015-DP0069).

\begin{figure}[h]
\centerline{
\epsfxsize=3in
\epsffile{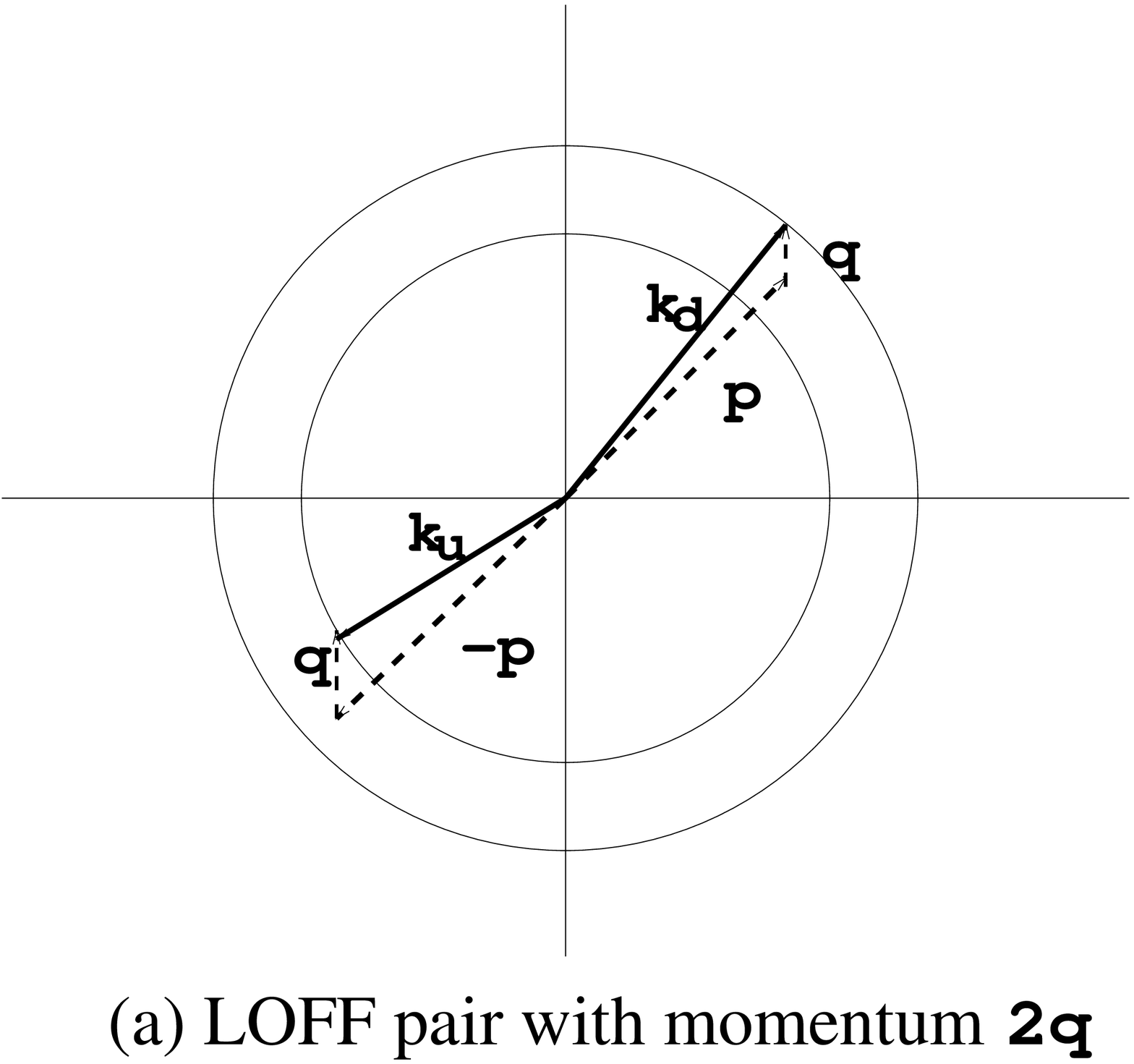}\hskip 0.2in
\epsfxsize=3in
\epsffile{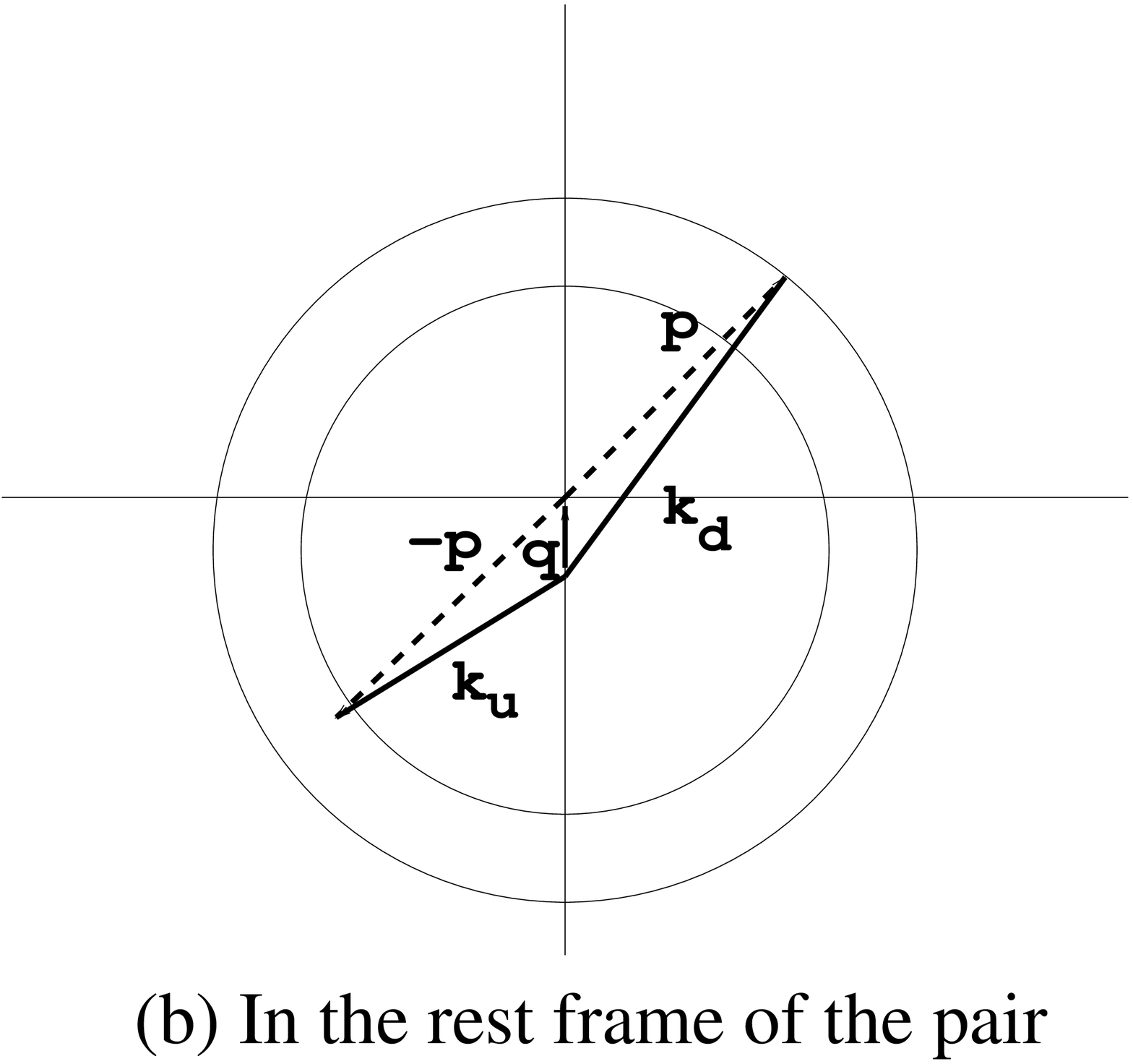}}
\vskip 0.2in
\caption[]{LOFF pairing of quarks with non-opposite momenta}
\label{loff1}
\end{figure}

\vfill


\end{document}